\documentstyle[preprint,epsf]{jpsj}

\title{Correlation of Firing in Layered Associative Neural Networks}

\author{Michiko Yamana and Masato Okada}

\author
{ 
Michiko {\sc Yamana}$^{1,}$\footnote{Present address:Central Research
Institute of Electric Power Industry, System Engineering Research
Laboratory, 2-11-1 Iwatokita, Komae-shi, Tokyo, 201-8511.}
and Masato {\sc Okada}$^{1,2}$
}

\inst
{
$^1$
RIKEN, Brain Science Institute, Laboratory for Mathematical Neuroscience,
2-1 Hirosawa, Wakou-shi, Saitama, 351-0198
\\
$^2$
Graduate School of Frontier Sciences, The University of Tokyo, 
Frontier Sciences Bldg 409, Kashiwanoha 5-1-5, Kashiwa-shi, Chiba,
277-8561\\
''Intelligent Cooperation and Control'', PRESTO, JST, Frontier
Sciences Bldg, Kashiwanoha 5-1-5, Kashiwa-shi, Chiba, 277-8561
}

\recdate
{
\today
}

\abst{
There is growing interest in a phenomenon called the ``synfire chain'',
in which firings of neurons propagate from pool to pool in the chain.
The mechanism of the synfire chain has been analyzed by many resarchers.
Keeping the synfire chain phenomenon in mind, we investigate a layered
associative memory neural network model, in which patterns are embedded
in connections between neurons. In this model, we also include
uniform noise in connections, which induces common input in the next layer.
Such common input in layers generate correlated firings of neurons.
We theoretically  obtain  the evolution of retrieval states in the case
of infinite pattern loading.
We find that the overlap between patterns and neuronal states is not
given as a deterministic quantity, but is described by a probability
distribution defined over the emsemble of synaptic matrices.
Our simulation results are in excellent agreement with
theoretical calculations.
}

\kword
{
correlation of firing, synfire chain, common input, associative memory model
}

\begin{document}
\sloppy
\maketitle
\section{Introduction}
Abeles proposed a synfire chain as a model of the cortex~\cite{abeles}.
It consists of neuron pools connected to each other in a
feedforward chain so that firings can propagate.
Repeated spatiotemporal firing patterns with precise
timing accuracy within milliseconds have been
recorded in the frontal cortex of monkeys performing tasks~\cite{abeles2}.
These findings suggest the occurence of some temporal structure in the brain.
In recent experiments,
temporally precise firing sequences of spikes have been reported
for a songbird vocal control system~\cite{prut,hahnloser}.
It was also proved that
sequences of synchronous activity propagate through a reproduced
multilayer feedforward network of neurons in an in vitro slice
preparation of rat cortex~\cite{kimpo,reyes,ikegaya}.

Models of a synfire chain and 
its mechanism have been studied theoretically by many
researchers~\cite{bienenstock,herrmann,hertz,diesmann,kato,amari,sakai,hamaguchi,hamaguchi2,kawamura}.
Recently, Diesmann and Cateau et al. investigated conditions for
stable propagation of correlated firings along 
the feedforward networks~\cite{diesmann,kato}.

Amari et al. analyzed a model of a neuron pool in which neurons receive
common overlapping input, and proved that higher order interactions of
neurons exist in synchronous firing, which generates  
wide-spread activity distributions.
That is, the self-averaging property breaks down even in the stochastic limit,
and activity depends on the sample, therefore activity distribution is
not concentrated on its mean value and  has a wide-spread distribution.
However, in their model, neurons are uniformly connected to each other and
each connection has no structures such as learning efficacies.

In the present study, 
we investigate a layered associative memory neural network model in which connections
patterns are embedded together with uniform noise which induces common input
into the next layer. 
Common input in layers generates correlated firing of neurons.
We discuss cases of infinite pattern loading in the networks.

\section{Layered Associative Memory Model}
We consider a layered associative memory model. 
In this model, neuron states are denoted by $x_i^\ell$ 
(neuron at $i$ site in layer $\ell$),
which takes $+1$ or $-1$ corresponding to the firing state and nonfiring state, respectively.
Firing of neurons is propagated according to
\begin{equation}
x_i^{\ell+1}=F \left( \sum_{j=1}^N J_{ij}^\ell x_j^\ell \right),
\label{eq:definition}
\end{equation}
where the layer number is $\ell=1,\cdots,L$ and the number of neurons in a layer
is $i=1,\cdots,N$. 
$F$ is the output function $F(h)= {\rm sgn}(h)$.
For the coupling of neurons, we consider the following form:
\begin{equation}
J_{ij}^\ell=\frac{1}{N}\sum_{\mu=1}^p \xi_i^{\ell+1,\mu} \xi_j^{\ell, \mu}+w_j^\ell .
\label{eq:J_ij}
\end{equation}
Here, the first term describes correlated learning in which the patterns are embedded.
Each component of the patterns is assumed to be an independent random variable
that takes a value of either $+1$ or $-1$ according to the probability,
\begin{equation}
{\rm Prob}[ \xi_i^{\ell,\mu} =\pm 1]=\frac{1}{2},~~~~~~ (\mu=1,\cdots,p,~\ell=0,\cdots,L),
\end{equation}
where the number of patterns is $p=\alpha N$.
The second term of Eq. (\ref{eq:J_ij}) is 
uniform Gaussian noise $w_j^\ell \sim {\cal N}(0,\delta^2/N)$.
The coupling $w_j^\ell$ propagates a 
firing of $x_j^\ell$ to the next layer uniformly, so it induces synfiring in the
next layer.

We introduce 
the order parameter characterizing memory retrieval, namely, 
the overlap with the $\mu$th pattern in the $\ell$th layer, defined by
\begin{equation}
m^{\ell,\mu}=\frac{1}{N}\sum_{i=1}^N \xi_i^{\ell,\mu} x_i^\ell .
\end{equation}

\section{Theory}
\label{sec:theory}

\subsection{Order parameter equations}
\label{ssec:order}
In the following analysis, we consider the situation where
the pattern $\mu=1$ is being retrieved, i.e., 
$m^{\ell,1} \equiv m^\ell \sim {\cal O}(1)$, 
and $m^{\ell,\mu \neq 1} \sim {\cal O}(1/\sqrt{N})$.
In this case, the input to the $\ell+1$th layer can be written as
\begin{equation}
h_i^\ell \equiv \sum_j J_{ij}^\ell x_i^\ell =\xi_i^{\ell+1}m^\ell +z_i^\ell+\eta^{\ell},
\label{eq:h_i}
\end{equation}
where
\begin{eqnarray}
&&z_i^\ell \equiv 
\frac{1}{N}\sum_j \sum_{\mu \ge 2} \xi_i^{\ell+1,\mu}\xi_j^{\ell,\mu} x_j^\ell ,
\label{eq:z_i}
\\
&&\eta^\ell \equiv \sum_j w_j^\ell x_j^\ell .
\end{eqnarray}
The first term of Eq. (\ref{eq:h_i})
is a signal term, the second term is the cross-talk noise term,
and the last term is the uniform noise term
$\eta^\ell \sim {\cal N}(0,\delta^2)$, which is independent of $i$.

The evolution of the neuron states then is given by 
\begin{equation}
x_i^{\ell+1}= F (h_i) =F \left(\xi_i^{\ell+1} m^\ell +z_i^\ell 
+\eta^\ell \right) .
\label{eq:x_eq}
\end{equation}
We note that in Eq. (\ref{eq:x_eq}), 
$\eta^\ell$ acts as a uniform threshold, so it induces synfiring in the $\ell+1$th layer.

Let us now derive equations describing the evolution of the order parameter $m^\ell$.
We can then follow the standard procedure~\cite{okada} to expand Eq. (\ref{eq:z_i}) as 
\begin{eqnarray}
&&z_i^{\ell+1}=\frac{1}{N} \sum_j \sum_{\mu \ge 2} \xi_i^{\ell+2,\mu} \xi_j^{\ell+1,\mu}
F \left[\xi_j^{\ell+1} m^\ell +\eta^\ell 
+\frac{1}{N} \sum_k \xi_j^{\ell+1,\mu} \xi_k^{\ell,\mu} x_k^\ell
+\frac{1}{N}\sum_{\nu \neq 1,\mu}\xi_j^{\ell+1,\nu} \xi_k^{\ell,\nu} x_k^\ell \right]
\nonumber
\\
&&\simeq \frac{1}{N} \sum_j \sum_{\mu \ge 2} \xi_i^{\ell+2,\mu} \xi_j^{\ell+1,\mu}
F \left[\xi_j^{\ell+1} m^\ell +\eta^\ell 
+\frac{1}{N}\sum_{\nu \neq 1,\mu}\xi_j^{\ell+1,\nu} \sum_k \xi_k^{\ell,\nu} x_k^\ell \right]
\nonumber
\\
&&+\frac{1}{N^2} \sum_j \sum_{\mu \ge 2} \sum_k \xi_i^{\ell+2,\mu} \xi_k^{\ell,\mu} x_k^\ell
F' \left[\xi_j^{\ell+1} m^\ell +\eta^\ell 
+\frac{1}{N}\sum_{\nu \neq 1,\mu}\xi_j^{\ell+1,\nu} \sum_k \xi_k^{\ell,\nu} x_k^\ell \right].
\end{eqnarray}
We then obtain 
\begin{eqnarray}
&&z_i^{\ell+1}=\frac{1}{N} \sum_j \sum_{\mu \ge 2} \xi_i^{\ell+2,\mu} \xi_j^{\ell+1,\mu}
x_j^{\ell+1,(\mu)}
\nonumber
\\
&&+ \sum_{n=0}^L \left( \prod_{t=0}^n U^{\ell+1-t} \right)
\frac{1}{N}\sum_j \sum_{\mu \ge 2} \xi_i^{\ell+2,\mu} \xi_j^{\ell-n,\mu} x_j^{\ell-n,(\mu)},
\end{eqnarray}
where we have introduced the following two quantities:
\begin{eqnarray}
&&x_j^{\ell+1,(\mu)}=F \left[\xi_j^{\ell+1} m^\ell +\eta^\ell 
+\frac{1}{N}\sum_{\nu \neq 1,\mu}\xi_j^{\ell+1,\nu} \sum_k \xi_k^{\ell,\nu} x_k^\ell \right]
\\
&&U^{\ell+1}=\frac{1}{N}\sum_j F'\left[\xi_j^{\ell+1} m^\ell +\eta^\ell 
+\frac{1}{N}\sum_{\nu \neq 1,\mu}\xi_j^{\ell+1,\nu} \sum_k \xi_k^{\ell,\nu} x_k^\ell \right].
\end{eqnarray}
As usual, we assume that $z_i^\ell$ obeys a Gaussian distribution,
where the mean equals $0$ and the variance of $z_i^\ell$,
${\rm E} [ (z_i^{\ell+1})^2 ]=(\sigma^\ell)^2$, is given by
\begin{eqnarray}
(\sigma^{\ell+1})^2 
&=&\alpha \left[ 1+(U^{\ell+1})^2(U^\ell)^2
+(U^{\ell+1})^2(U^\ell)^2(U^{\ell-1})^2 +\cdots \right]
\nonumber
\\
&=&\alpha \left[ 1+(U^{\ell+1})^2 \frac{(\sigma^\ell)^2}{\alpha} \right]
\nonumber
\\
&=&\alpha+(U^{\ell+1})^2 (\sigma^\ell)^2.
\label{eq:sigma}
\end{eqnarray}
Finally, when $m^\ell, \sigma^\ell$ and $\eta^\ell$ are given,
we obtain the following coupled order parameter equations
using $F(h)={\rm sgn}(h)$:
\begin{eqnarray}
 m^{\ell+1}(m^\ell, \sigma^\ell ,\eta^\ell )
&=& \int Dz^\ell \langle \xi^{\ell+1} F 
(\xi^{\ell+1} m^\ell +z^\ell+\eta^\ell) \rangle_\xi
\nonumber
\\
&=&  \frac{1}{2}\left[ {\rm erf}(u)+{\rm erf}(v)\right],
\label{eq:m}
\\
U^{\ell+1}(m^\ell, \sigma^\ell ,\eta^\ell )
&=&\ \int Dz^\ell \langle F' 
(\xi^{\ell+1} m^\ell +z^\ell+\eta^\ell) \rangle_\xi
\nonumber
\\
&=& \frac{1}{\sqrt{2 \pi}\sigma^\ell}
\left( {\rm e}^{-u^2}+{\rm e}^{-v^2} \right),
\label{eq:u}
\\
Dz^\ell &=&\frac{dz^\ell}{\sqrt{2 \pi} \sigma^\ell} 
\exp \left(-\frac{(z^\ell)^2}{2(\sigma^\ell)^2} \right),
\nonumber
\end{eqnarray}
where 
$u=(m^\ell+\eta^\ell)/(\sqrt{2}\sigma^\ell),v=(m^\ell-\eta^\ell)/(\sqrt{2}\sigma^\ell)$
and erf$(x)=\frac{2}{\sqrt{\pi}}\int_0^x {\rm e}^{-y^2} dy$.
$\langle ~~\rangle_\xi$ denotes the average over random patterns of one layer.
Here, we use ${\rm sgn}'(x)=2 \delta(x)$.

\subsection{Probability distribution function}
\label{ssec:prob}
Let us now discuss the effect of $\eta$ theoretically.
In the presence of $\eta^\ell$, the evolution of $x_i^\ell$ is given by Eq.~(\ref{eq:x_eq}).
For a given $\eta^\ell$, 
the order parameters at the $\ell+1$th layer are given in terms of those
at the $\ell$th layer as Eq.(\ref{eq:sigma})-(\ref{eq:u})
However, $\eta^\ell$ is distributed in each layer $\ell$ according to
the Gaussian distribution,
and thus the evolution is described by the distribution function $p(m^\ell, \sigma^\ell)$.
In general, the distribution is described by the three parameters as 
$p(m^\ell, \sigma^\ell,\eta^\ell)$.
In this case, they can be factorized as $p(m^\ell, \sigma^\ell)p(\eta^\ell)$
and thus the evolution of the distribution $p(m^\ell, \sigma^\ell)$
can be described by
\begin{eqnarray}
&&p(m^{\ell+1}, \sigma^{\ell+1})=\int dm^\ell \int d\sigma^\ell \int d\eta^\ell
p(m^\ell, \sigma^\ell)p(\eta^\ell)
\\
\nonumber
&&\times \delta(m^{\ell+1}-m^{\ell+1}(m^\ell,\sigma^\ell,\eta^\ell))
\delta(\sigma^{\ell+1}-\sigma^{\ell+1}(m^\ell,\sigma^\ell,\eta^\ell)),
\end{eqnarray}
where $\delta(\cdot)$ is Dirac's delta function and
$
p(\eta^\ell) = \frac{1}{\sqrt{2 \pi} \delta^\ell }
\exp \left(-\frac{(\eta^\ell)^2}{2(\delta^\ell)^2} \right).
$

If we set
\begin{eqnarray}
&& m^{\ell+1} (m^\ell, \sigma^\ell , \eta^\ell) \equiv f (m^\ell,
\sigma^\ell , \eta^\ell) = \frac{1}{2}\left[ {\rm erf}(u)+{\rm erf}(v)
\right],
\label{eq:f}
\\
&& \sigma^{\ell+1} (m^\ell, \sigma^\ell , \eta^\ell) \equiv g(m^\ell,
\sigma^\ell , \eta^\ell) = \sqrt{ \alpha +\frac{1}{2 \pi} ( {\rm
e}^{-u^2}+ {\rm e}^{-v^2} ) },
\label{eq:g}
\end{eqnarray}
transforming $m^\ell, \sigma^\ell$ into $f(m^\ell, \sigma^\ell,
\eta^\ell), g(m^\ell, \sigma^\ell, \eta^\ell)$, 
we get the evoution equation of the distribution,
\begin{equation}
p(m^{\ell+1}, \sigma^{\ell+1})=
\int_{-\infty}^{\infty} \frac{ d \eta^\ell p(\eta^\ell) p(m^{\ell *},
\sigma^{\ell *} ) }
{ \left| \frac{\partial f}{\partial m^\ell} \frac{\partial g}{\partial
\sigma^\ell } -
\frac{\partial g}{\partial m^\ell } \frac{\partial f}{\partial
\sigma^\ell }  
\right|_{m^\ell=m^{\ell *}, \sigma^\ell=\sigma^{\ell *}} },
\end{equation}
where $m^{\ell *}, \sigma^{\ell *}$ satisfy eqs.(\ref{eq:f}) and
(\ref{eq:g}).
Then we obtain the probability distribution function as
\begin{equation}
p(m^{\ell+1}, \sigma^{\ell+1})
=\int_{-\infty}^\infty d\eta^\ell p(\eta^\ell) 
\frac{2\pi(\eta^\ell)^2 \sqrt{ 2\pi \alpha+\left( {\rm e}^{-(u^*)^2}+
 {\rm e}^{-(v^*)^2} \right)^2} }
{(u^*-v^*)^4 {\rm e}^{-(u^*)^2-(v^*)^2} \left( {\rm e}^{-(u^*)^2}+{\rm e}^{-(v^*)^2} 
\right) } p(m^{\ell *}, \sigma^{\ell *} ),
\end{equation}
where $u^*=(m^{\ell *}+\eta^{\ell})/( \sqrt{2} \sigma^{\ell *} )$
and $v^*=(m^{\ell *}-\eta^{\ell})/( \sqrt{2} \sigma^{\ell *} )$.


\section{Results}
\label{sec:results}
\begin{figure}[htb]
\begin{center}
\leavevmode
\epsfxsize=90mm
\epsfbox{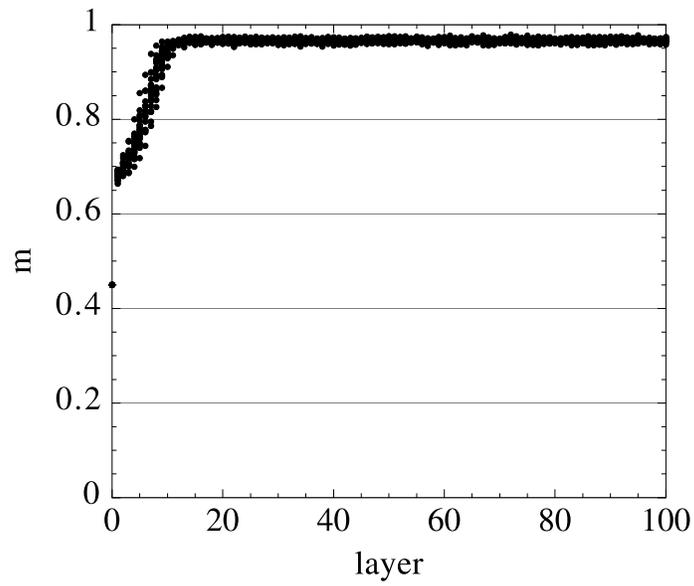}
\end{center}
\caption{Simulation results for the evolution of $m^\ell$ for
$\alpha=0.2, \delta=0, N=10000, m^0=0.45$. We plot the results for $20$ samples.
}
\label{Fig:1}
\end{figure}

\begin{figure}[htb]
\begin{center}
\leavevmode
\epsfxsize=90mm
\epsfbox{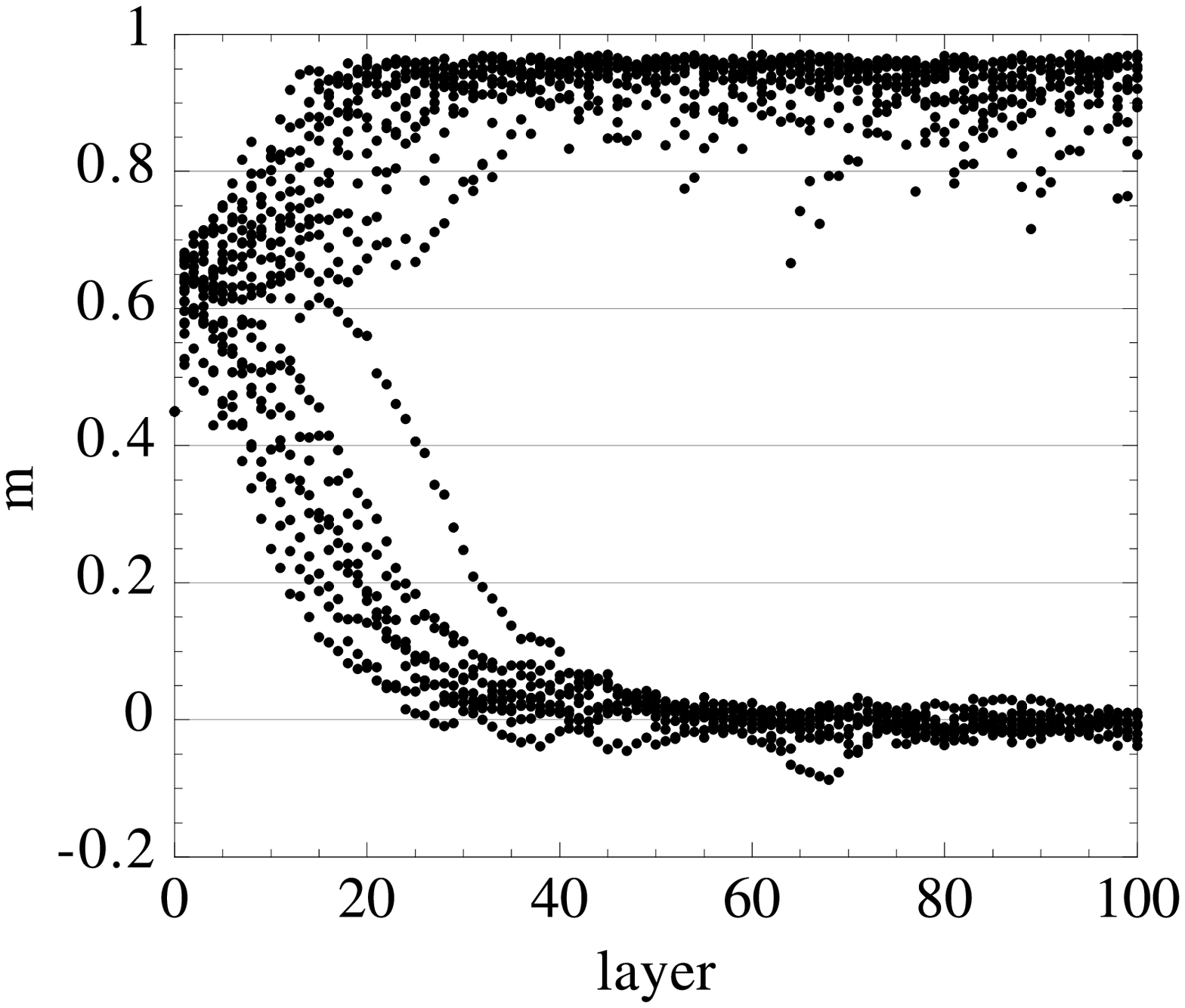}
\end{center}
\caption{Simulation results for the evolution of $m^\ell$ for
$\alpha=0.2, \delta=0.2, N=10000, m^0=0.45$. We plot the results for $20$ samples.
}
\label{Fig:2}
\end{figure}

\begin{figure}[htb]
\begin{center}
\leavevmode
\epsfxsize=90mm
\epsfbox{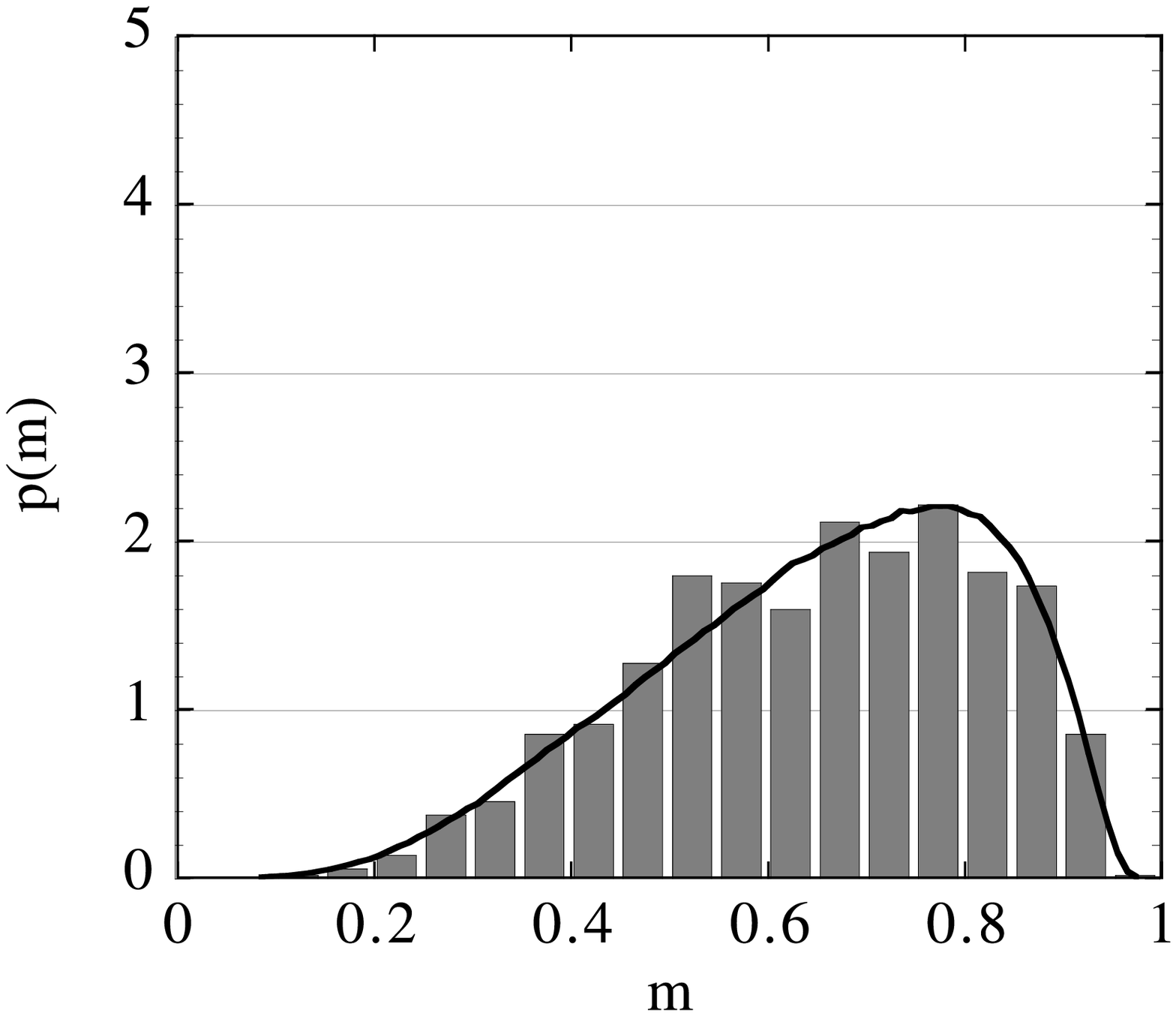}
\end{center}
\caption{Layer=10.  
The histogram is simulation results ($N=10000, 1000$samples).
The line is a theoretical calculation of the distribution $p(m)$.
}
\label{Fig:3}
\end{figure}

\begin{figure}[htb]
\begin{center}
\leavevmode
\epsfxsize=90mm
\epsfbox{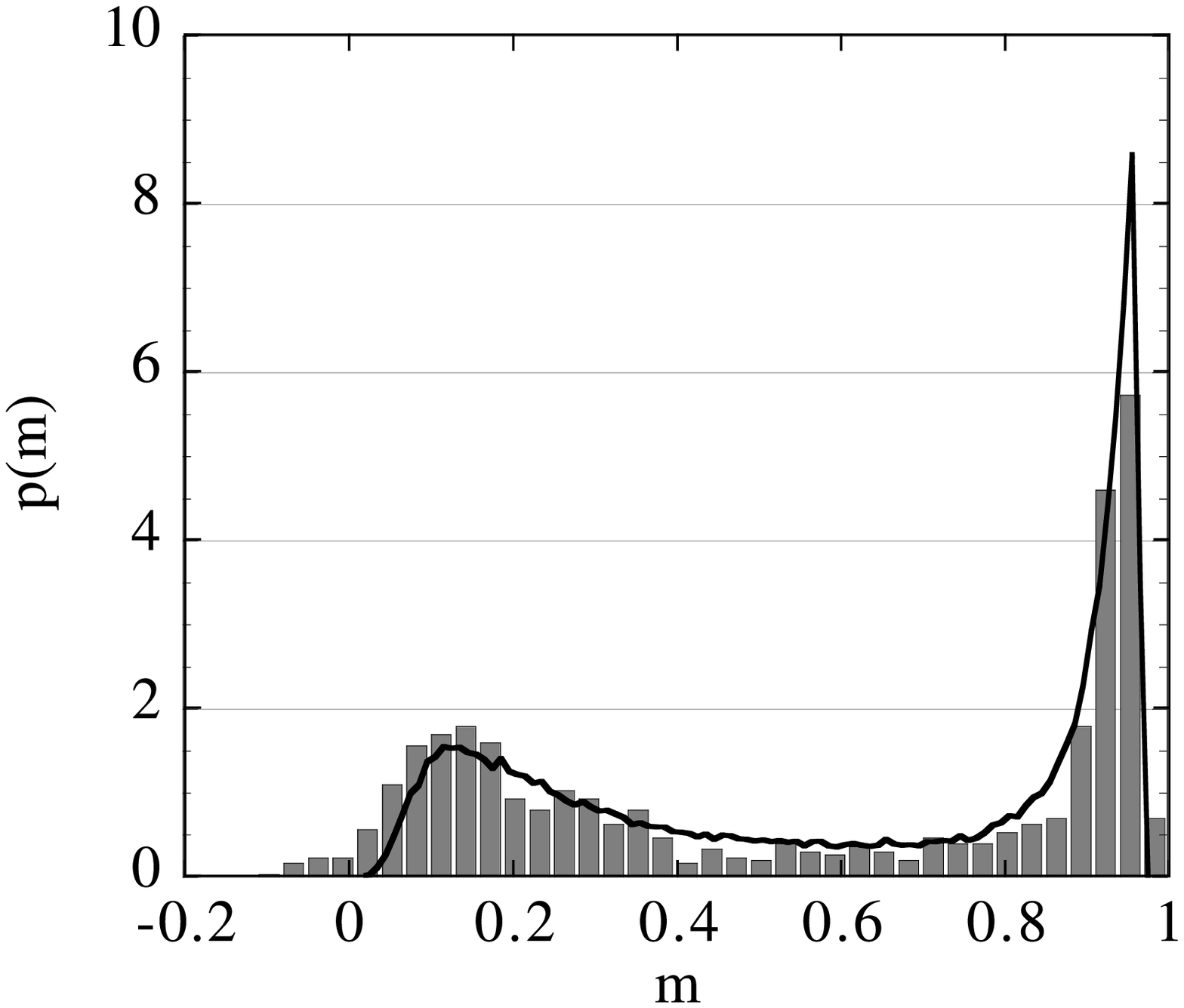}
\end{center}
\caption{Layer=20.
The histogram is simulation results ($N=10000, 1000$samples).
The line is a theoretical calculation of the distribution $p(m)$.
}
\label{Fig:4}
\end{figure}

\begin{figure}[htb]
\begin{center}
\leavevmode
\epsfxsize=90mm
\epsfbox{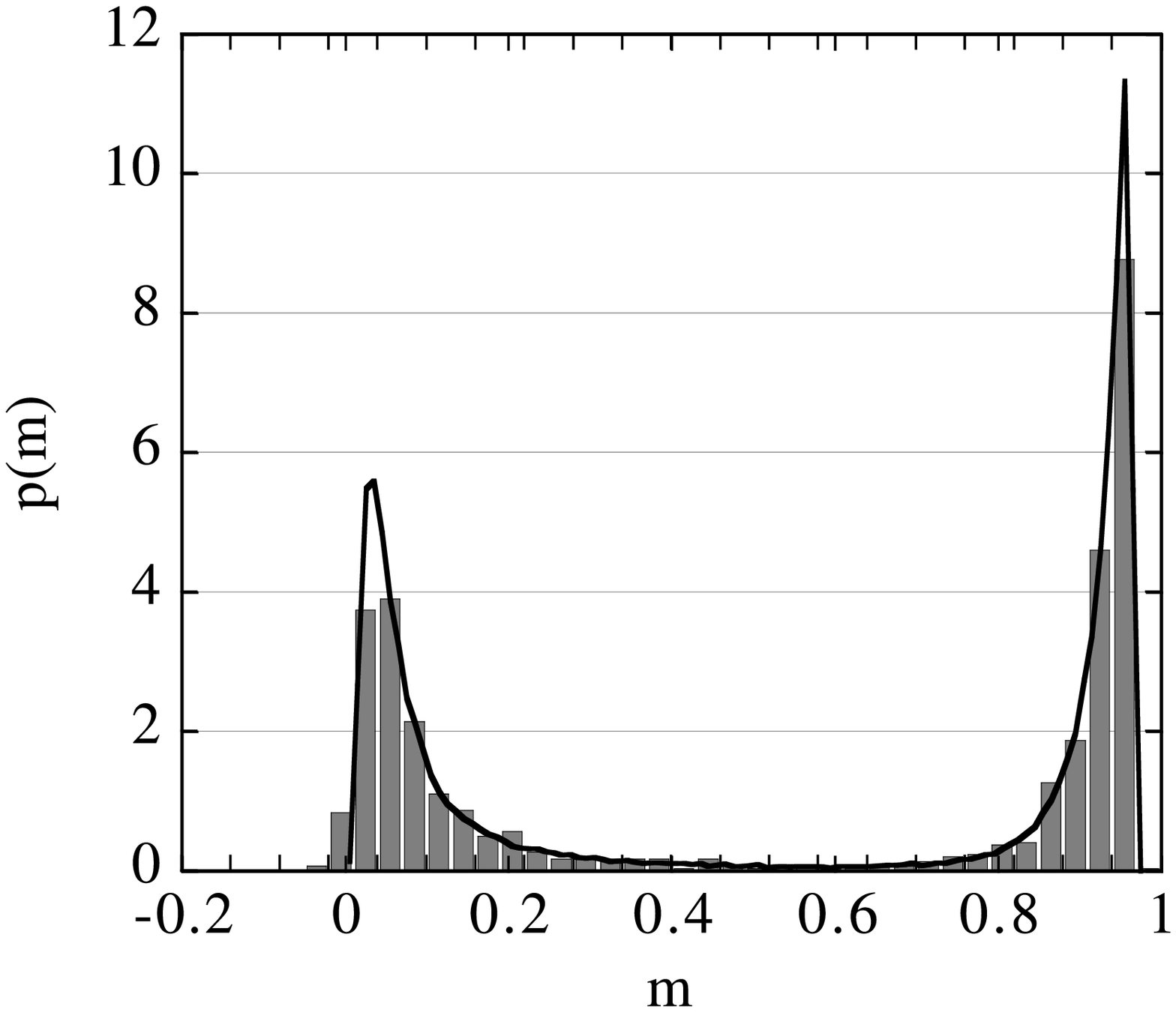}
\end{center}
\caption{Layer=30.
The histogram is simulation results ($N=10000, 1000$samples).
The line is a theoretical calculation of the distribution $p(m)$.
}
\label{Fig:5}
\end{figure}

\begin{figure}[htb]
\begin{center}
\leavevmode
\epsfxsize=90mm
\epsfbox{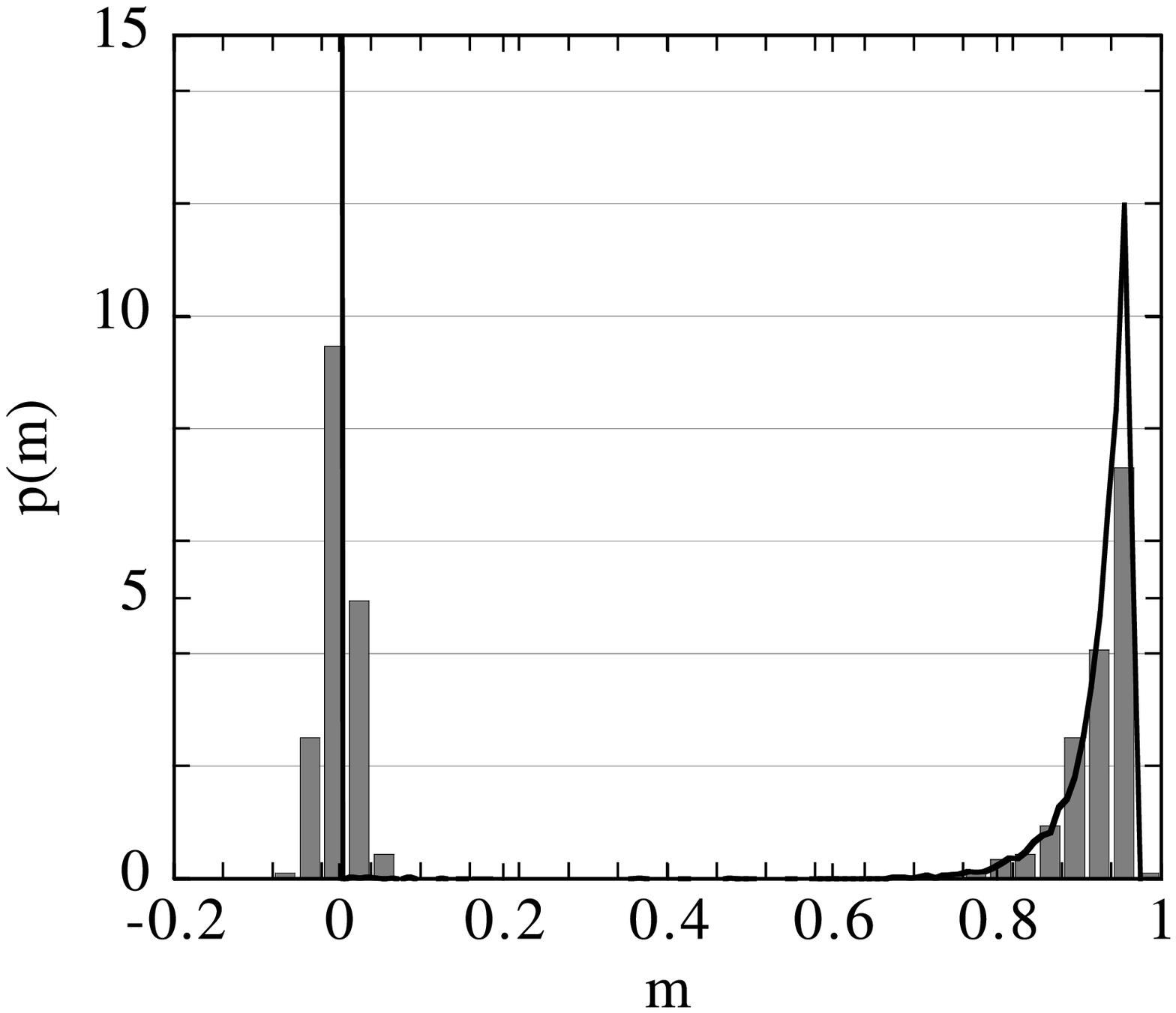}
\end{center}
\caption{Layer=100.
The histogram is simulation results ($N=10000, 1000$samples).
The solid line is a theoretical calculation of the distribution $p(m)$.
}
\label{Fig:6}
\end{figure}

We show the simulation results of Eq. (\ref{eq:definition}) 
to see the effects of $\eta$.
As an initial condition,
we set the state of the first layer according to the following probability:
\begin{equation}
{\rm Prob}[x_i^0=\pm 1] =\frac{1 \pm m^0 \xi_i^0}{2} .
\end{equation}
Figure \ref{Fig:1} shows the evolution of the overlap $m$ in the case of no uniform
Gaussian noise ($\delta=0$).
In this case, the evolution is deterministic.
Figure \ref{Fig:2} shows the simulation results including noise $\eta$.
In this case, the evolution is not deterministic, but distributed.
In Figures \ref{Fig:3} - \ref{Fig:6}, 
we plot the distribution function $p(m)$ for the four layers
($\ell=10, 20, 30, 100$), where $p(m)=\int d\sigma p(m,\sigma).$
The histogram shows simulation results ($N=10000, 1000$ samples).
The solid lines are theoretical calculations of the distribution $p(m)$.
We find excellent agreement between simulation results and calculations.
According to the evolution of layers,
one can see that the distribution is peaked at two states,
namely, the nonretrieval state and the retrieval state.
It is noted that the nonretrieval state does not exist in the finite
loading case.

\begin{figure}[htb]
\begin{center}
\leavevmode
\epsfxsize=100mm
\epsfbox{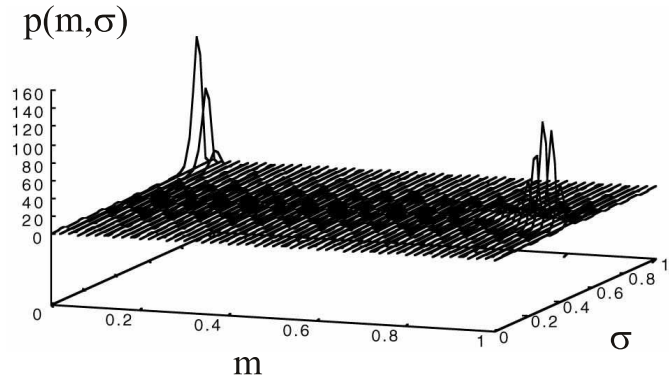}
\end{center}
\caption{$p(m,\sigma)$ ($\alpha=0.2,\delta=0.2,N=200000,m^0=0.45, \ell=100$).
}
\label{Fig:7}
\end{figure}

\begin{figure}[htb]
\begin{center}
\leavevmode
\epsfxsize=90mm
\epsfbox{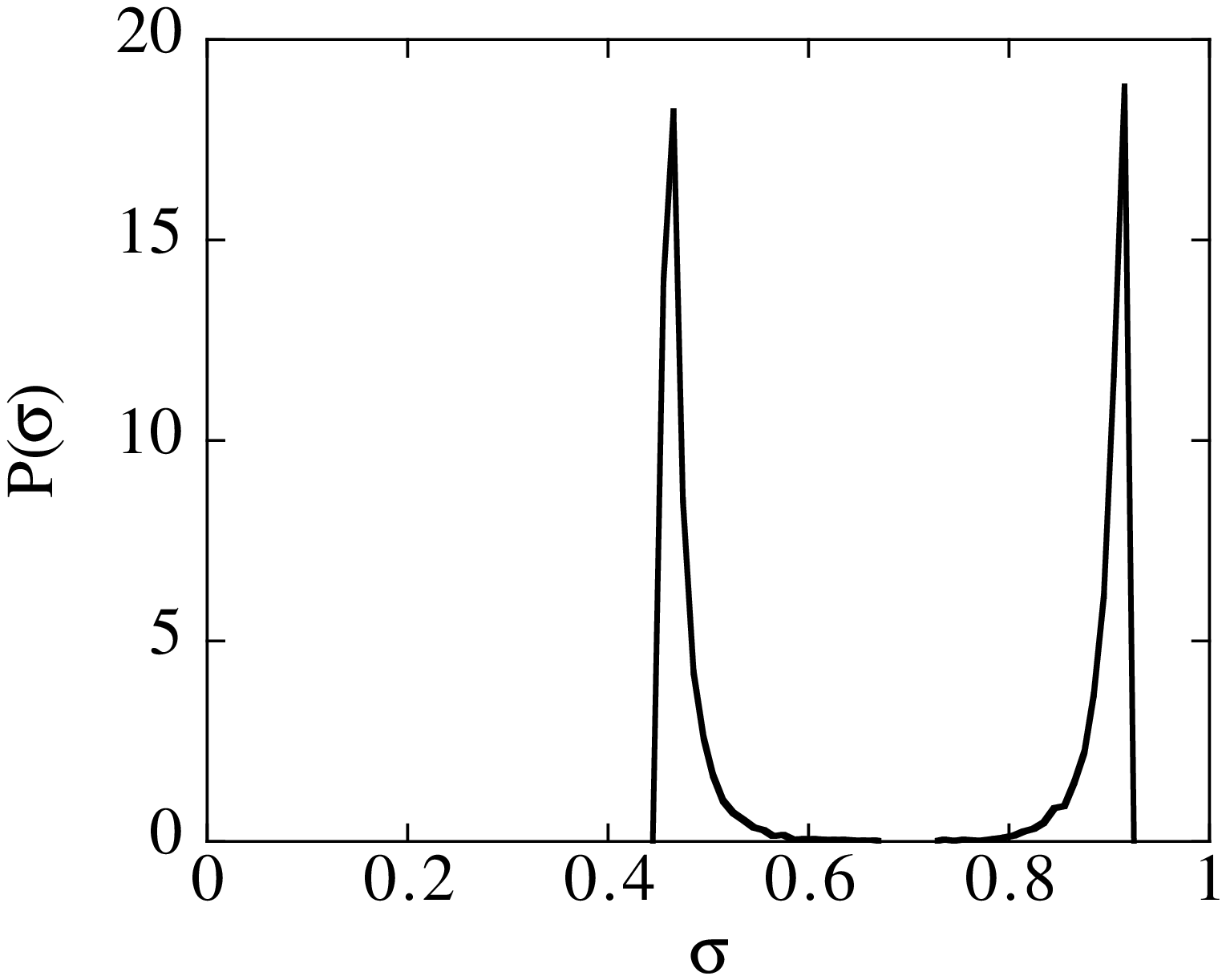}
\end{center}
\caption{$p(\sigma)$.  Integrating $p(m,\sigma)$ over $m$.
}
\label{Fig:8}
\end{figure}

Figure \ref{Fig:7} shows the distribution function
$p(m,\sigma)$ ($\alpha=0.2,\delta=0.2,m^0=0.45, \ell=100$).
One peak with $m$ close to zero corresponds to the nonretrieval state.
The other peak with $m$ close to one corresponds to the retrieval state.
Integrating $p(m,\sigma)$ over $m$, we obtain $p(\sigma)$.
Two peaks corresponding to the retrieval state and the nonretrieval state 
can be seen in Fig. \ref{Fig:8}.

\section{Summary}
We investigated the layered associative memory neural network model,
in which patterns are embedded in connections between neurons. In this
model, we also include
uniform noise in connections, which induces correlated firings of
neurons.
We theoretically  obtain  the evolution of retrieval states in the case
of infinite pattern loading.
We find that the overlap between patterns and neuronal states is not
given as a deterministic quantity, but is described by a probability
distribution.
Our simulations results are in excellent agreement with
theoretical calculations.





\begin{thebibliography}{99}



\bibitem{abeles}M. Abeles:
{\it Corticonics}, (Cambridge Univ. Press, Cambridge, 1991).

\bibitem{abeles2}M. Abeles, H. Bergman, E. Margalit and E. Vaadia:
J. Neurophysiol. {\bf 70} (1993) 1629.

\bibitem{prut}Y. Prut, E. Vaadia, H. Bergman, I. Haalman, H. Slovin
and M. Abeles:
J. Neurophysiol. {\bf 79} (1998) 2857.

\bibitem{hahnloser}H. Hahnloser, A. Kozhevnikov and M. Fee:
Nature {\bf 419} (2002) 65.

\bibitem{kimpo}R. Kimpo, F. Theunissen and A. Doupe:
J. Neurosci. {\bf 23} (2003) 5750.

\bibitem{reyes}A. Reyes:
Nature Neurosci. {\bf 6} (2003) 593. 

\bibitem{ikegaya}Y. Ikegaya, G. Aaron, R. Cossart, D. Aronov,
I. Lample, D. Ferster and R.Yuste:
Science {\bf 304} (2004) 559.

\bibitem{bienenstock}E. Bienenstock:
Network: Computat. Neural Syst. {\bf 6} (1995) 179.

\bibitem{herrmann}M. Herrmann, J. Hertz and A. Pr\"{u}gel-Bennett:
Network: Computat. Neural Syst. {\bf 6} (1995) 403.

\bibitem{hertz}J. Hertz and A. Pr\"{u}gel-Bennett:
Network: Computat. Neural Syst. {\bf 7} (1996) 357.


\bibitem{diesmann}M. Diesmann, M-O. Gewaltig and AD Aertsen:
Nature {\bf 402} (1999) 529.

\bibitem{kato}H. Cateau and T. Fukai:
Neural Networks {\bf 14} (2001) 675.

\bibitem{amari}S. Amari, H. Nakahara, S. Wu and Y. Sakai:
Neural Comp. {\bf 15} (2003) 127.

\bibitem{sakai}Y. Sakai:
private communication.

\bibitem{hamaguchi}K. Hamaguchi, M. Okada, M. Yamana and K. Aihara:
IEICE Trans, D-II {\bf J87} (2004) 1689

\bibitem{hamaguchi2}K. Hamaguchi, M. Okada, M. Yamana and K. Aihara:
Neural Comp, to be published.

\bibitem{kawamura}M. Kawamura, M. Yamana and M. Okada:
Tech. Rep. of IEICE, NC2003 {\bf 103} (2004) 127.

\bibitem{okada}M. Okada:
Neural Networks {\bf 8} (1995) 833.

\end{thebibliography}
\end{document}